\newcommand{\eq}[2]{\begin{equation}{#1}\label{#2}\end{equation}}
\def\natural{{\mathchoice
    {\hbox{ $\displaystyle\kern-1.4mm 1\kern-.7mm {\rm N}$}}
    {\hbox{ $\textstyle\kern-1.4mm 1\kern-.7mm {\rm N}$}}
    {\hbox{$\scriptstyle\kern-1.4mm 1\kern-.7mm {\rm N}$}}
    {\hbox{$\scriptscriptstyle\kern-1.4mm 1\kern-.7mm {\rm N}$}}}}
\def\real{{\mathchoice
    {\hbox{$\displaystyle\kern-.2mm 1\kern-.8mm {\rm R}\kern-.2mm$}}
    {\hbox{$\textstyle\kern-.2mm 1\kern-.8mm {\rm R}\kern-.2mm$}}
    {\hbox{$\scriptstyle\kern-.2mm 1\kern-.8mm {\rm R}\kern-.2mm$}}
    {\hbox{$\scriptscriptstyle \kern-.2mm 1\kern-.8mm {\rm R}\kern-.2mm$}}}}
\def\complex{{\mathchoice
    {\hbox{$\displaystyle\kern-.2mm {\rm C}\kern-1.5mm\raise.2mm
			       \hbox{\vrule height6pt}\kern1.3mm$}}
    {\hbox{$\textstyle\kern-.2mm {\rm C}\kern-1.5mm\raise.3mm
			       \hbox{\vrule height6pt}\kern1.3mm$}}
    {\hbox{$\scriptstyle\kern-.2mm{\rm C}\kern-1.5mm\raise.2mm
			       \hbox{\vrule height3pt}\kern1.3mm$}}
    {\hbox{$\scriptscriptstyle\kern-.2mm{\rm C}\kern-1.5mm\raise.2mm
			      \hbox{\vrule height2pt}\kern1.3mm$}}}}
\def\summ#1#2#3{\sum\limits_{#1}^{#2}\lower3pt\hbox{${ }_{#3}$}}
\def\row#1#2{({#1}_1,\ldots,{#1}_{#2})}

\documentstyle[11pt]{article}
\normalbaselines
\begin{document}
\bibliographystyle{unsrt}

\vbox{\vspace{38mm}}
\begin{center}
{\LARGE \bf
Four Dimensional Orbit Spaces of \\[2mm]
Compact Coregular Linear Groups}\\[5mm]

G. Sartori\\{\it
Dipartimento di Fisica, Universit\`a di Padova and INFN, Sezione di Padova\\
I--35131 Padova, Italy (e-mail: gfsartori@padova.infn.it)}\\[3mm]
V. Talamini \\
{\it INFN, Sezione di Padova\\
I--35131 Padova, Italy (e-mail: talamini@padova.infn.it)}\\[5mm]
{(Submitted \ \ October 1993)}\\[5mm] \end{center}

\begin{abstract}

All four dimensional orbit spaces of compact coregular linear groups have been
determined.  The results are obtained through the integration of a universal
differential equation, that only requires as input the number of elements of
an integrity basis of the ideal of polynomial invariants of the linear group.
Our results are relevant and lead to universality properties in the physics of
spontaneous symmetry breaking at the classical level.

\end{abstract}

\section{Introduction}

In theories in which the ground state of the system is determined by a minimum
of a potential $V(x),\ x\in\real^n$, which is invariant under the
transformations of a compact group $G$ acting linearly in $\real^n$, the
characterization of the schemes of spontaneous symmetry breaking rests on the
determination of the minimum of the potential.	Owing to the high number of
variables involved and the degeneracy of the minimum along an orbit of $G$,
this is generally a difficult problem to solve, even if a polynomial
approximation is used for the potential.

The degeneracy associated with the G-invariance of the potential can be
eliminated by considering the potential as a function in the orbit space
$\real^n/G$ (see ref.\cite{683} and references therein).  If $V(x)$ is a
polynomial
or a $C^\infty$-function, this is easily achieved in the following way.  Let
$\{p_1(x), \dots , p_q(x)\}$ be a minimal integrity basis (hereafter
abbreviated in IB) of the ring $\real^n[x]^G$ of polynomial invariants of $G$.
The IB defines an {\it orbit map} $p(x)$, mapping $\real^n$ onto a
semi-algebraic subset, $p(\real^n)$, of $\real^q$, and $p(\real^n)$ is a
faithful image of $\real^n/G$ (see, for instance, \cite{610} or \cite{710}
and references therein) whose geometric stratification \cite{840} is strictly
related to the isotropy type stratification of the orbit space \cite{080}.
The invariance properties of the potential allow to write it in the form $\hat
V(p(x))$ \cite{285,617,690}.  The function $\hat V(p)$, which is defined in
the whole of $\real^q$, when restricted to $p(\real^n)$ has the same
regularity properties and the same range as $V(x)$.  Thus it can be
advantageously used in the determination of the minimum of the potential \cite
{278,313,316,020,021}.

As long as the potential is not specified, each point $p\in p(\real^n)$ can be
considered as the representative of a possible ground state of the system.
Points lying on the same stratum represent possible ground states of the
system and their invariance groups are conjugated subgroups of $G$.  The set
$p(\real^n)$ yields therefore a suggestive geometric picture of the possible
configurations (phases) of the system after spontaneous symmetry breaking.

Often, invariance properties are the only bounds which are imposed to the
potential, beyond regularity and stability properties and/or bounds to the
degree, when the potential is a polynomial.  If the symmetry groups of the
potentials of different theories share isomorphic orbit spaces, the potentials
have the same formal expression if they are written as functions in orbit
space, despite the completely different physical meaning of the variables and
parameters involved in their definitions.  Thus, the problems of determining
the geometric features of the phase space, the location and the stability
properties of the minima of the potential, the number of phases and the
allowed phase transitions, reduce to identical mathematical problems in all
these theories \cite{021,681,682}.  This remark makes the orbit space approach
to a model independent analysis of spontaneous symmetry breaking particularly
appealing.

The price to pay in the orbit space approach to the minimization of a
potential $V$ is essentially twofold:

\begin{enumerate}
\item IB's are sometimes difficult to determine.
\item The domain of the function $\hat V(p)$
is not the whole of $\real^q$, but reduces to the semi-algebraic subset
$p(\real^n)$ \cite{020,021}, which is not trivial to determine.
\end{enumerate}

The difficulties mentioned under items 1 and 2 can however be bypassed.  In
fact, it has been shown \cite{020,021,651}, that the polynomial relations
defining $p(\real^n)$
and its strata can be determined from the positivity properties of a matrix
$\hat P(p)$, defined only in terms of the gradients of the
elements of an IB.  The semialgebraic subset of $\real^q$ where $\hat P(p)\ge
0$ and has rank $k$ is the union of all the $k$-dimensional strata.  If $G$ is
a compact {\it coregular} linear group (hereafter abbreviated in CCLG), i.e.\
if there is no algebraic relation among the elements of an IB, the matrix
$\hat P(p)$ plays the role of an inverse metric matrix in the interior of
$\real^n/G$ and the isomorphism classes of the orbit spaces of all the compact
coregular linear groups can be classified in terms of equivalence classes of
matrices $\hat P(p)$.  The matrices $\hat P(p)$ have been shown \cite{681,682}
to be solutions of a canonical differential equation in the variables
$p_1,\dots ,p_q$, that contains as free parameters only the degrees $d_1,\dots
,d_q$ of the IB.  Therefore, they can all be determined, {\it even if the
classification of the compact coregular linear groups is not yet complete
and/or the explicit form of the elements $p_a(x)$ of the corresponding IB's is
not known}.

For $q\le 3$, all the relevant solutions of the canonical equation have been
provided in a preceding paper \cite{682}; they share the following remarkable
features that have been conjectured to hold for all values of $q$:

\begin{description}
\item[a)]
Only particular values of the ratios between the degrees are allowed and
for each choice of the degrees, there is only a finite
number of non equivalent solutions.
\item[b)] In convenient IB's, the coefficients of the polynomials $\hat
P_{ab}(p)$ are integer numbers.
\end{description}

In this paper, after recalling the basic steps leading to the formulation of
the canonical equation, we shall provide all its solutions in the case of four
dimensional orbit spaces.  From the explicit form of the solutions it will be
evident that the statements listed under items {\bf a--b} remain true for
$q=4$ too.

The plan of the paper is the following.  In Section 2 we shall recall some
known results concerning the characterization of orbit spaces (see, for
instance, \cite{683,710,080,021})
and in Sections 3 and 4 we shall resume the main results of reference
\cite{682}.  In Section 5 the explicit expressions of the relevant
solutions of the canonical equation for $q=4$ are reported.

\section{Orbit Spaces of Compact Linear Groups}

Let $G$ be a compact group of $n\times n$ matrices acting linearly in the
Euclidean space $\real^n$.  We shall assume, without loss of generality, that
$G$ is a subgroup of $SO_n(\real)$.

If $G_x$ denotes the isotropy subgroup of $G$ at $x$, the class of the
isotropy subgroups at the points of the orbit $\Omega $ through $x$, is the
class of subgroups conjugated to $G_x$ in $G$ and is called the {\it orbit
type} of $\Omega $.  All the points lying on orbits with the same orbit type
form an {\it isotropy type stratum of} ({\it the action of $G$ in}) $\real^n$.
Isotropy type strata are in a one-to-one correspondence with orbit types and
the connected components of a stratum are smooth manifolds with the same
dimensions.

The orbit space of the action of $G$ in $\real^n$ is the quotient space
$\real^n/G$, endowed with the quotient topology and differentiable structure.
The connected components of the image in orbit space of an isotropy type
stratum of $\real^n$ are smooth manifolds with the same dimensions.

Almost all the images of the orbits of $G$ lie in a unique stratum $\Sigma_p$
of $\real^n/G$, the {\it principal stratum}, which is a connected open dense
subset of $\real^n/G$.	The boundary $\bar\Sigma_p\backslash\Sigma_p$ of the
principal stratum is the union of disjoint {\it singular} strata.  All the
strata lying on the boundary $\bar\Sigma\backslash\Sigma$ of a stratum
$\Sigma$ are open in $\bar\Sigma\backslash\Sigma$.

The collection of all orbit types is a finite and partially ordered set:  the
orbit type of a stratum $\Sigma$ is contained in the orbit types of the strata
lying in its boundary and there is a unique minimum orbit type, the {\it
principal orbit type}, corresponding to the principal stratum.

A faithful image of the orbit space $\real^n/G$ and its stratification can be
obtained from classical invariant theory in the following way.	Let
$\{p_1(x), \dots , p_q(x)\}$ denote an homogeneous IB of the ring
$\real^n[x]^G$ of all $G$-invariant polynomials and $d_i$ denote the
homogeneity degree of $p_i(x)$.

The {\it orbit map}, $p :\ \real^n\rightarrow \real^q :\ x\rightarrow (p_1(x),
\dots ,p_q(x)),$ maps all the points of $\real^n$ lying on an orbit of $G$
onto
a unique point of $\real^q$ and induces a diffeomorphism of $\real^n/G$ onto a
{\it semi-algebraic q-dimensional connected closed} subset $p(\real^n)$ of
$\real^q$.  Like all semialgebraic sets, $p(\real^n)$ is the disjoint union of
finitely many connected semialgebraic differentiable varieties $\{E_i\}$ ({\it
the primary strata}), such that the boundary of each $E_i$ is empty or the
union of lower dimensional primary strata and each primary stratum is open in
its closure.  The connected components of the isotropy type strata of
$\real^n/G$ are in a one-to-one correspondence with the primary strata of
$p(\real^n)$; the interior of $p(\real^n)$ represents the principal stratum,
the boundary hosts the singular strata.

There is some arbitrariness in the choice of the elements of an IB, but their
number $q$ and their degrees are determined by $G$ alone.  If $G$ is {\it
coregular}, as we shall assume in the rest of the paper, $p(\real^n)$ is a
$q$-dimensional subset of $\real^q$.

If $\{p\}$ and $\{p'\}$ are distinct IB's of $G$, then $p(\real^n)$ and
$p'(\real^n)$ are isomorphic semi-algebraic sets, since the IB transformation
$p'(p)$ is a bi-rational mapping.

\section{The matrix $\hat P(p)$}

The classification of the orbit spaces of all CCLG's can be reduced to the
classification of the orbit spaces of the CCLG's with no fixed points, as we
shall do.  In fact, if in $\real^n$ there are linear $G$--invariants,
generating the linear subspace $W$ of $\real^n$, then $\real^n/G$ is
homeomorphic to $W\times W_\bot/G_\bot$, where the linear group $G_\bot$
denotes the reduction of the linear group $G$ to the orthogonal complement
$W_\bot $ of $W$ in $\real^n$.	In these cases the degrees of all the
polynomial invariants are necessarily $\ge 2$ and the following conventions
can be adopted:

\eq{p_q(x)=\summ 1ni x_i^2.}{(2.2a)}
\eq{d_1\ge d_2\ge\dots \ge d_q=2,}{(2.2b)}
Hereafter, by an IB we shall always mean a {\it minimal homogeneous integrity
basis} fulfilling eqs.(\ref {(2.2a)}) and (\ref {(2.2b)}).

A polynomial $\hat F(p)$ defines a $G$-invariant polynomial $F(x)=\hat
F(p(x))$.  Vice-versa, for each invariant polynomial $F(x)$, there exist a
unique polynomial $\hat F(p)$, such that $F(x)=\hat F(p(x))$ \cite{285,617}.
$\hat F(p)$ will be said to be $w$-homogeneous with weight $w$ if $\hat
F(p(x))$ is homogeneous in $x$ with degree $w$.

The set $p(\real^n)$ can be characterized by means of a $q\times q$ matrix
$\hat P(p)$, defined in $\real^q$ by the following relations \cite{020,021}:

\eq{P_{ab}(x)=\summ 1ni\partial_i p_a(x)\partial_i p_b(x)=
\hat P_{ab}(p(x)),\qquad a,b=1,\dots ,q.}{(2.10)}

The matrix $\hat P(p)$ has the following properties:

\begin{description}
\item[P1.] {\it Symmetry}.  The matrix $\hat P(p)$ is symmetric.

\item[P2.] {\it Homogeneity}.  The matrix elements $\hat
P_{ab}(p)$ are real $w$-homogeneous polynomial functions of $p$ and
\eq{w(\hat P_{ab})=d_a+d_b-2;}{(3.1a)}
moreover, owing to eq. (\ref{(2.2a)})
\eq{\hat P_{qa}(p)=\hat P_{aq}(p)=2d_ap_a,\qquad a=1,\dots ,q.}{(3.1b)}

\item[P3.]  {\it Tensor character.} $\hat P(p)$ transforms as a
rank 2 contravariant tensor under IB transformations (hereafter abbreviated in
IBT's) $p'=f(p),$ where $f(p)$ is a convenient $w$-homogeneous polynomial map,
whose Jacobian matrix $J(p)$ is an upper block-triangular matrix with constant
diagonal blocks, \eq{\hat P'(p'(p))=J(p) \hat P(p) J^T(p),}{(2.11)} The
determinant of $\hat P(p)$ is a relative invariant of the group of IBT's.

\item[P4.] {\it Positivity}.  The set $p(\real^n)$ is the only subset of
$\real^q$ where $\hat P(p)\ge 0$.  In the interior of $p(\real^n)$, the rank
of $\hat P(p)$ equals $q$; on the boundary, it is lower and almost everywhere
equal to $q-1$.  The subset of $\real^q$ where $\hat P(p)\ge 0$ and has rank
$k$ is the union of all the $k$-dimensional strata \cite{020,021,651}.	The
image $\bar{\cal S}=p(S^{n-1})$ of the unit sphere of $\real^n$ is a compact
connected $(q-1)$-dimensional semialgebraic subset of the hyperplane $\Pi
=\{p\in \real^q\mid p_q=1\}$, which is the intersection of $p(\real^n)$ with
$\Pi$ \cite{681,682}.

\item[P5.] {\it Boundary conditions} \cite{681,682}.  Let ${\cal I}(\hat\sigma
)$ denote the ideal of all the polynomials in $p$ vanishing on $\hat\sigma$, a
$(q-1)$-dimensional primary stratum (or a union of primary strata) of
$p(\real^n)$.  Then, for all $f\in {\cal I}(\hat\sigma )$ \eq{\summ 1qb\hat
P_{ab}(p)\partial_b f(p)\in{\cal I}(\hat\sigma ).}{112}
\end{description}

\vskip2truemm Two $\hat P$-matrices, related by a relation like
eq.(\ref{(2.11)}) will be said to be {\it equivalent}.	The $q$-dimensional
orbit spaces of the CCLG's $G$ and $G'$, acting in $\real^n$ and,
respectively, $\real^{n'}$, will be said to be {\it isomorphic} if their $\hat
P$ matrices are equivalent; in particular, in case of isomorphic orbit spaces,
$p'(\real^{n'})=f(p(\real^n))$, where $p'=f(p)$ has the form of an IBT.

The identification of the class of all IB's with the class of coordinate
systems makes the interior of $\real^n/G$ a differentiable manifold and $\hat
P(p)$ yields an inverse metric matrix in it.  Thus, in order to classify the
isomorphism classes of the orbit spaces of all the CCLG's, it is
sufficient to classify the equivalence classes of matrices $\hat P(p)$.

\section{The Canonical Equation}

The results summarized below have been proved in \cite{681,682}.

If $\hat{\cal B}$ is the boundary of $p(\real^n)$, the ideal ${\cal
I}(\hat{\cal B})$, formed by the polynomials in $p$ vanishing in the whole
of $\hat{\cal B}$, has a unique independent generator $A(p)$.
For $f(p)=A(p)$, eq.(\ref{112}) reduces to

\eq{\summ 1qb\hat P_{ab}(p){\partial\over\partial p_b}A(p)=\lambda^{(A)}_a(p)
A(p), \qquad a=1,\dots ,q,}{(3.3)}
where $\lambda^{(A)}(p)$ is a contravariant vector field with
$w$-homogeneous components.

The vector $\lambda^{(A)}(p)$ can be reduced to the canonical form
$\lambda_a^{(A)}(p)=2\delta _{aq}w(A)$ in the so-called {\it A-bases}.	In an
$A$-basis, the boundary conditions assume the following {\it canonical} form:

\eq{\summ 1qb \hat P_{ab}(p)\partial_b A(p) = 2\delta _{aq}w(A)A(p),\qquad
a=1,\dots ,q.}{a1}

{}From eq.(\ref{a1}) one deduces that, in every $A$-basis, the following facts
are true:

\vskip3truemm\begin{description}
\item[i)] $A(p)$ is a factor of $\det \hat P(p)$ and its sign can be chosen to
be positive inside $p(\real^n)$; we shall call it a {\it complete active
factor}. Consequently:
\eq{w(A)\le w(\det\hat P)=2\summ 1qa d_a -2q.}{a2a}
\item[ii)] $A(p)\big |_{p_q=1}$ has a unique local non degenerate maximum
lying at $p_0=(0,\dots , 1)$.  This means that $p_0$ is in the interior of
$\bar{\cal S}$ and that the Hessian matrix $H(p)$ of $A(p)$ is negative
definite at $p_0$.  This allows, in particular, to set a lower bound to the
weight of $A$:	\eq{w(A)\ge 2d_1.}{a2b}

\item[iii)] $\hat P(p_0)$ is proportional to $H(p_0)$;
it is block diagonal and, in a subclass of $A$-bases ({\it standard
$A$-bases}), it is diagonal:
\eq{\hat P_{ab}(p_0)=d_ad_b\delta_{ab}, \qquad a,b=1, \dots, q.}{a2c}
Two different standard $A$-bases are related by an IBT
$p'_\alpha =f_\alpha (p),\ \alpha = 1, \dots , q-1,$ not involving $p_q$; the
corresponding Jacobian matrix is orthogonal at $p_0$.
\end{description}

\vskip3mm In eq.(\ref{a1}), $\hat P_{ab}(p)$ and $A(p)$ can be considered
as unknown polynomials satisfying only the conditions listed under items
{\bf P1-P3}. In this case
eq.({\ref{a1}) will be called the {\it canonical equation}.

Let us now assume that the only information we have on a CCLG with no fixed
points is the number of elements of its IB's.  Then, the $\hat P$-matrix of
$G$, in a standard $A$ basis, has to be searched among the solutions of the
canonical equation (in which the degrees $d_i$ are considered as integer
parameters $\ge 2$), satisfying the {\it initial conditions} specified in
eq.(\ref{a2c}).  We shall call these solutions {\it allowable $\hat P$
matrices}.

\vskip3mm If the degrees $d_a$ are given explicit numerical values, the
polynomials $\hat P_{ab}(p)$ and $A(p)$ can be expanded in the $p_a$'s.  After
elimination of these variables, the canonical equation reduces to a system of
algebraic equations in the unknown coefficients of the polynomials.  For $q=3$
the number of variables and equations is already frightening.

If the degrees are maintained as free parameters, as we did, only the
dependence on $p_1$ can be rendered explicit in the unknown polynomials
involved in the canonical equation.  After the elimination of $p_1$, the
canonical equation expands into a seemingly awfull system of coupled algebraic
and differential equations.  For $q\le 4$, a clever use of the initial
conditions has allowed to solve it.  A complete list of the allowable
solutions for $q=3$ has been published in ref.\cite{682}; a list of the
allowable solutions for $q=4$ will be given in the next section.

The allowable solutions we have found for $q=3$ and for $q=4$ share the
following relevant features:

\vskip3truemm
\begin{description}
\item[a)] The canonical equation admits allowable solutions only for
particular values of the degrees and for each choice of the
degrees there is only a finite number of non equivalent (with respect to
IBT's) allowable solutions.

\vskip3mm\item[b)] Let $D$ denote the MCD of $d_1,\dots ,d_{q-1}$ and $R$ the
following $q\times q$ matrix, defined in $\Pi$:
\eq{R_{ab}(p_1,\dots ,p_{q-1})=\hat P_{ab}(p_1,\dots ,p_{q-1}
,1)/(d_a d_b).}{u1}
For each allowable solution corresponding to the weights $(d_1,\dots
,d_{q-1},2)$, there is tower of allowable solutions corresponding to the
weights $(s d_1/D,\dots ,s d_{q-1}/D,2)$, where $s\in\natural$ is a scale
parameter.  Different solutions belonging to the same tower are not
equivalent, but share the same $R$-matrix; as a consequence, their positivity
regions, in the hyperplane $\Pi$, are isomorphic.

\vskip3mm\item[c)] The $A$-basis can be chosen so that all the coefficients of
the polynomials $\hat P_{ab}(p)$ are integer numbers.

\vskip3mm\item[d)] All the allowable solutions satisfy the following
positivity condition, analogous to the positivity condition stated under item
{\bf P4}:  {\it $\hat P(p)$ is $\ge 0$ in a unique connected semi-algebraic
subset of $\real^q$, whose intersection with $\Pi$ is a $(q-1)$-dimensional,
compact, connected semi-algebraic subset, $\bar{\cal S}$, of $\real^{q-1}$.}

\vskip3mm\item[e)] $A(p)$ is a factor of $\det\hat P(p)$ which vanishes on the
whole boundary of the region where $\hat P(p)\ge 0$ and there
are no other factors of $\det \hat P(p)$ vanishing in a $(q-1)$-dimensional
semi-algebraic subset of the boundary.
\end{description}

\vskip3mm The statement under item {\bf b)}, the fact that
$A(p)$ is a factor of $\det\hat P(p)$ (see item {\bf e)}) and the
fact that $\hat P(p)|_{p_q=1}$ is positive semi-definite {\it only} in a
compact set (see item {\bf d)}), hold true for all values of $q$
\cite{682,998}.  Even if we have no proof that also the other features of the
allowable solutions we have listed are independent of the particular range of
values chosen for $q$, we believe they yield a reasonable support to the
following conjecture \cite{681}:

\vskip5mm \proclaim Conjecture.  For all integer $q>1$ the allowable solutions
of the canonical equation fullfil the conditions stated above under items {\bf
a), c)}, {\bf d)} and {\bf e)}.

\vskip5mm We have not proved that every allowable solution of the canonical
equation yields the $\hat P$-matrix of a CCLG; it is clear, however, that the
$\hat P$-matrix of every CCLG is in the set of these solutions.  Therefore,
the selection rules mentioned under item {\bf a)} are obeyed also by the
CCLG's; moreover, the orbit spaces of all the CCLG's, whose IB's are formed by
the same number $q$ ($\le 4$, and, if the conjecture holds true, for all $q$)
of elements with the same degrees $\row dq$, can be classified in a finite
number of isomorphism classes; for each CCLG, the IB can be chosen so that the
$\hat P$-matrix elements are $w$-homogeneous polynomials with integer
coefficients.

\vskip3mm
All the finite coregular linear groups have been classified and the
corresponding IB's determined \cite{731}.  All the compact coregular simple
real linear Lie groups, and in many cases their IB's, can be determined from
the results obtained by Schwarz for the complex coregular simple linear Lie
groups \cite{711}.  The $\hat P$-matrices of the images of the 2-, 3-
\cite{683} and 4-dimensional \cite{000} orbit spaces of all these real groups
should be found among the allowable solutions of the canonical equation we
have found.  For some of them the identification is immediate, since there is
only one allowable solution with suitable degrees, for the other cases a
direct check is needed \cite{000}.  At our knowledge, in the literature there
is no complete classification of general compact coregular linear Lie groups.

\vskip5mm Some of the solutions we have found may correspond also to {\it
non-}coregular CLG's.  A criterium to recognize such solutions will be
presented in a forthcoming paper \cite{000}.

\section{Allowable solutions of the canonical equation for $q=4$}

Below we shall report a (hopefully) complete list of the matrices $R$
characterizing the distinct towers of allowable solutions of the canonical
equation for $q=4$.  The representative matrices $R(p_1,\dots ,p_{q-1})$ are
expressed in a convenient $A$-basis, $A$ being a complete active factor of
$\det \hat P$, with weight $d_A$.  The corresponding matrices $\hat
P(p_1,\dots ,p_q)$ can be easily recovered from eq.(\ref{u1}) and the
$w$-homogeneity of $\hat P_{ab}(p)$.  We shall also make use of the
following definition:

\eq{\det R =\rho A.}{uuu}

The towers of allowable solutions will be labelled by an indexed capital latin
letter and a set of positive integer parameters $j_1,\ j_2,\dots
$\footnote{The list of allowable solutions given in ref.\cite{999} is
incomplete and the labels used to distinguish the different towers do
not generally agree with the labels used in the present paper.}
Different labels generally correspond to unequivalent allowable
$\hat P$-matrices. The degrees are expressed as functions of $s$ and of the
$j_i$'s.  The limitations on the possible values of $s$ and of the $j_i$'s
coming from eq.(2) will be understood.
\vfill\eject

\vglue10truemm\centerline{\bf Class $A1(j_1,j_2,j_3,j_4)$}
\nobreak\vskip5truemm\nobreak
\noindent $d_A=2 d_1,\quad
d_1=s({j_1}+{j_2})({j_3}+{j_4})/4,\quad d_2=s({j_1}+{j_2})/2,\quad d_3=s;$

\nobreak\vskip3truemm\nobreak\noindent
$j_1\ge {j_2}$ and $j_1={j_2}$ for odd $d_3;\quad $ ${j_3}\ge {j_4}$
and ${j_3}={j_4}$ for odd $j_1$ or ${j_2};$

\nobreak\vskip3truemm\nobreak\noindent $R_{11}=
[j_1 j_2+(j_1-j_2)p_3] [j_3 j_4 (j_1-p_3)^{j_1/2}(j_2+p_3)^{j_2/2}+(j_3-j_4)
p_2] [j_3(j_1-p_3)^{j_1/2}(j_2+p_3)^{j_2/2}-p_2]^{j_3-1}[j_4  (j_1-p_3)^{j_1/2}
(j_2+p_3)^{j_2/2}+ p_2]^{j_4-1}(j_1-p_3)^{j_1/2-1}(j_2+p_3)^{j_2/2-1} ,$

\nobreak\vskip3truemm\nobreak\noindent $R_{12}=0,$

\nobreak\vskip3truemm\nobreak\noindent $R_{13}=0,$

\nobreak\vskip3truemm\nobreak\noindent $R_{22}=[j_1j_2+(j_1-j_2)p_3] [j_3j_4
(j_1-p_3)^{j_1/2}(j_2+p_3)^{j_2/2}+(j_3-j_4)p_2 ]
(j_1-p_3)^{j_1/2-1}(j_2+p_3)^{j_2/2-1} ,$

\nobreak\vskip3truemm\nobreak\noindent $R_{23}=0,$

\nobreak\vskip3truemm\nobreak\noindent $R_{33}= j_1j_2+(j_1-j_2)p_3,$

\nobreak\vskip3truemm\nobreak\noindent $A=[j_3 (j_1-p_3)^{j_1/2}(j_2+p_3)^
{j_2/2} -p_2 ]^{j_3} [j_4 (j_1-p_3)^{j_1/2}(j_2+p_3)^{{j_2}/2} +p_2]^{j_4}-
p_1^2.$

\nobreak\vskip3truemm\nobreak\noindent $\rho = [j_1j_2 +(j_1-j_2)p_3]^2
[j_3j_4 (j_1-p_3)^{j_1/2}(j_2+p_3)^{j_2/2} +(j_3-j_4)p_2]
(j_1-p_3)^{j_1/2-1}(j_2+p_3)^{j_2/2-1} .$

\vskip22truemm\centerline{\bf Class $A2(j_1,j_2,j_3,j_4)$}
\nobreak\vskip5truemm\nobreak
\noindent $d_A=2 d_1,\quad d_1=s[({j_1}+1)({j_2}+{j_3})/2+{j_4}],\quad
d_2=s({j_1}+1), \quad d_3=2s;$

\nobreak\vskip3truemm\nobreak\noindent ${j_3}\ge {j_2},$ and ${j_3}={j_2}$ for
even ${j_1};$
\quad $j=(j_1+1)(j_2+j_3);$

\nobreak\vskip3truemm\nobreak
\noindent $R_{11}=\{(j+2j_4) [2jj_4-(j-2j_4)p_3] [j_2j_3 (p_3+j)^{(j_1+1)/2}+
 (j_2-j_3)p_2] (p_3+j)^{(j_1 -1)/2}-4j_4^2p_2^2 \} (2j_4-p_3)^{j_4-1}
 [j_2 (p_3+j)^{(j_1+1)/2}-p_2]^{j_2-1} [j_3 (p_3+j)^{(j_1+1)/2}+p_2]^{j_3-1},$

\nobreak\vskip3truemm\nobreak\noindent $R_{12}=0,$

\nobreak\vskip3truemm\nobreak\noindent $R_{13}=0,$

\nobreak\vskip3truemm\nobreak\noindent $R_{22}=(j+2 j_4) [j_2 j_3 (p_3+j)^
{(j_1+1)/2}+ (j_2-j_3)p_2](p_3+j)^{(j_1-1)/2},$

\nobreak\vskip3truemm\nobreak\noindent $R_{23}=2 {j_4} p_2,$

\nobreak\vskip3truemm\nobreak\noindent $R_{33}=2 j{j_4}- (j-2 j_4)p_3,$

\nobreak\vskip3truemm\nobreak\noindent $A=(j+2j_4) (2 {j_4}-p_3)^{j_4} [{j_2}
(p_3+j)^{({j_1}+1)/2}- p_2]^{j_2} [{j_3} (p_3+j)^{({j_1}+1)/2}+p_2]^{j_3}-
p_1^2,$

\nobreak\vskip3truemm\nobreak\noindent $\rho =(j+2 {j_4})
[2 j j_4-(j-2 j_4)p_3][j_2j_3 (p_3+j)^ {(j_1+1)/2}+(j_2-j_3)p_2]
(p_3+j)^{(j_1-1)/2}-4 j_4^2 p_2^2.$

\vfill\eject\vglue10truemm\centerline{\bf Class $A3(j_1,j_2,j_3)$}
\nobreak\vskip5truemm\nobreak
\noindent $d_A=2 d_1+d_3,\quad
d_1={s} (j_1+1) ({j_2}+{j_3})/2,\quad d_2={s} (j_1+1) ,\quad d_3=2s;$

\nobreak\vskip3truemm\nobreak\noindent ${j_2}\ge {j_3},$ and ${j_2}={j_3}$ for
even $j_1;$
\quad $j=(j_1+1)(j_2+j_3)/2;$

\nobreak\vskip3truemm\nobreak\noindent $
R_{11}=(j+1)[{j_2} {j_3} (p_3+j)^{(j_1+1)/2}+ ({j_3}-{j_2})p_2]
[{j_2} (p_3+j)^{(j_1+1)/2}+p_2]^{{j_2}-1}
[{j_3} (p_3+j)^{(j_1+1)/2}-p_2]^{{j_3}-1}
(p_3+j)^{(j_1-1)/2},$

\nobreak\vskip3truemm\nobreak\noindent $ R_{12}=0,$

\nobreak\vskip3truemm\nobreak\noindent $ R_{13}=p_1,$

\nobreak\vskip3truemm\nobreak\noindent $
R_{22}=(j+1) [{j_2} {j_3} (p_3+j)^{(j_1+1)/2}+ ({j_3}-{j_2})p_2]
(p_3+j)^{(j_1-1)/2},$

\nobreak\vskip3truemm\nobreak\noindent $ R_{23}=p_2,$

\nobreak\vskip3truemm\nobreak\noindent $ R_{33}=j+(1-j)p_3,$

\nobreak\vskip3truemm\nobreak\noindent $
A=(1-p_3)\{[{j_2} (p_3 +j)^{(j_1+1)/2}+p_2]^{j_2}[{j_3}
(p_3+j)^{(j_1+1)/2}-p_2]^{j_3} - p_1^2\},$

\nobreak\vskip3truemm\nobreak\noindent $ \rho =(j+1)^2 [{j_2} {j_3} (p_3+j)^
{(j_1+1)/2}+
({j_3} -{j_2})p_2](p_3+j)^{(j_1-1)/2}. $

\vskip22truemm\centerline{\bf Class $A4(j_1,j_2)$}
\nobreak\vskip5truemm\nobreak
\noindent $d_A=2 d_1+{j_1}d_3,\quad d_1=2 {s} {j_2},\quad
d_2={s} ({j_1}+{j_2}), \quad d_3=2 {s};$

\nobreak\vskip3truemm\nobreak\noindent $
R_{11}=({j_1}+{j_2})({j_1}+2 {j_2})[2 {j_2}({j_1}+{j_2})+p_3]^{2 {j_2}-1},$

\nobreak\vskip3truemm\nobreak\noindent $
R_{12}= j_2 ({j_1}+2 {j_2})p_2 [2 {j_2}({j_1}+{j_2})+p_3]^{j_2-1} $

\nobreak\vskip3truemm\nobreak\noindent $ R_{13}=j_1(j_1+j_2) p_1,$

\nobreak\vskip3truemm\nobreak\noindent $
R_{22}=(j_1+2j_2)[j_1(j_1+j_2)-p_3]^{j_1-1}\{j_1^2 p_1+
[2 {j_1} {j_2} ({j_1 }+{j_2})^2+({j_1}^2-2{j_2}^2)p_3]
[2{j_2}({j_1}+{j_2})+p_3]^{j_2-1}\},$

\nobreak\vskip3truemm\nobreak\noindent $ R_{23}= -{j_1} {j_2} p_2,$

\nobreak\vskip3truemm\nobreak\noindent $ R_{33}=({j_1}+{j_2})[2j_1{j_2}({j_1}+
{j_2})+ ({j_1}-2j_2)p_3],$

\nobreak\vskip3truemm\nobreak\noindent $
A=\{[2 {j_2}({j_1}+{j_2})+p_3]^{j_2}-p_1\}\{({j_1}+{j_2})
[{j_1}({j_1}+{j_2})-p_3]^{j_1}\big [[2j_2({j_1}+{j_2})+p_3]^{j_2}+p_1\big ]
-p_2^2\},$

\nobreak\vskip3truemm\nobreak\noindent $ \rho
= ({j_1}+2 {j_2})^2\{j_1^2p_1+[2j_1j_2(j_1+j_2)^2+({j_1}^2-2{j_2}^2)p_3]
[2 {j_2}({j_1}+{j_2})+p_3]^{j_2-1}\}. $

\vfill\eject\vglue10truemm\centerline{\bf Class $A5(j_1,j_2,j_3)$}
\nobreak\vskip5truemm\nobreak
\noindent $d_A=2 d_1+d_2,\quad d_1=s(j_1j_2+j_3),
\quad d_2=2sj_2 ,\quad d_3=2 {s} ;$

\nobreak\vskip3truemm\nobreak\noindent $j=j_1j_2+j_2,\qquad j'=j_1j_2+j_3;$

\nobreak\vskip3truemm\nobreak\noindent $
R_{11}=(j+j_3)j'(j'j_3-p_3)^{j_3-1}\{{j_1}
[j{j'}^2j_3+(j_3^2-jj_1j_2)p_3] (jj'+p_3)^{j_2-1}+
 j_3^2p_2\} [j_1 (jj'+p_3)^{j_2}+p_2]^{{j_1}-1},$

\nobreak\vskip3truemm\nobreak\noindent $
R_{12}=j_1j_2 (j+j_3) p_1 (jj'+p_3)^{j_2-1},$

\nobreak\vskip3truemm\nobreak\noindent $ R_{13}=-j_2j_3p_1,$

\nobreak\vskip3truemm\nobreak\noindent $
R_{22}=j'(j+{j_3}) (jj'+p_3)^{j_2-1} [j_1(jj'+p_3)^{j_2}+(1-j_1)p_2],$

\nobreak\vskip3truemm\nobreak\noindent $ R_{23}=j'j_3 p_2,$

\nobreak\vskip3truemm\nobreak\noindent $
R_{33}=j'[jj'j_3+(j_3-j)p_3],$

\nobreak\vskip3truemm\nobreak\noindent $
A=[(jj'+p_3)^{j_2}-p_2] \{{j'}^2 (j'j_3-p_3)^{j_3}
[j_1(jj'+p_3)^{j_2}+p_2]^{j_1}-p_1^2\},$

\nobreak\vskip3truemm\nobreak\noindent $ \rho
=(j+{j_3})^2\{j_1[j{j'}^2j_3-(jj_1j_2-j_3^2)p_3]
(jj'+p_3)^{j_2-1} +j_3^2p_2\}.$

\vskip22truemm\centerline{\bf Class $A6(j_1,j_2,j_3)$}
\nobreak\vskip5truemm\nobreak
\noindent $d_A=2 d_1+d_2,\quad d_1={s} ({j_1}+{j_2}) ({j_3}+1)/2 ,
\quad d_2={s} ({j_1}+{j_2}), \quad d_3={s};$

\nobreak\vskip3truemm\nobreak\noindent ${j_1}\ge {j_2}$ and $j_2=j_1$ for odd
$d_3.$

\nobreak\vskip3truemm\nobreak\noindent $
R_{11}=(j_3+2)[{j_1}{j_2}+({j_1}-{j_2})p_3]
({j_1}-p_3)^{{j_1}-1} ({j_2}+p_3)^{{j_2}-1}
[({j_3}+1)({j_1}-p_3)^{j_1}({j_2}+p_3)^{j_2}+p_2]^{j_3}
,$

\nobreak\vskip3truemm\nobreak\noindent $
R_{12}=p_1 [{j_1}{j_2}+({j_1}-{j_2})p_3]
 ({j_1}-p_3)^{{j_1}-1}({j_2}+p_3)^{{j_2}-1},$

\nobreak\vskip3truemm\nobreak\noindent $ R_{13}=0,$

\nobreak\vskip3truemm\nobreak\noindent $
R_{22}=[{j_1}{j_2}+({j_1}-{j_2})p_3]
 ({j_1}-p_3)^{{j_1}-1}({j_2}+p_3)^{{j_2}-1}
[({j_3}+1)({j_1}-p_3)^{j_1}({j_2}+p_3)^{j_2}-{j_3}p_2],$

\nobreak\vskip3truemm\nobreak\noindent $ R_{23}=0,$

\nobreak\vskip3truemm\nobreak\noindent $ R_{33}={j_1}{j_2}+({j_1}-{j_2})p_3,$

\nobreak\vskip3truemm\nobreak\noindent $
A=[ ({j_1}-p_3)^{j_1}({j_2}+p_3)^{j_2}-p_2] \{[({j_3}+1)
 ({j_1}-p _3)^{j_1}({j_2}+p_3)^{j_2}+p_2]^{j_3+1}
-p_1^2\},$

\nobreak\vskip3truemm\nobreak\noindent $ \rho
= (j_3+2)[{j_1}{j_2}+({j_1}-{j_2})p_3]^2
 ({j_1}-p_3)^{{j_1}- 1}({j_2}+p_3)^{{j_2} - 1}.$

\vfill\eject\vglue10truemm\centerline{\bf Class $A7(j_1,j_2)$}
\nobreak\vskip5truemm\nobreak
\noindent $d_A=2 d_1+d_2+d_3,\quad d_1={s} {j_1} ({j_2}+1),\quad
d_2=2 {s} {j_1} , \quad d_3=2 {s};$

\nobreak\vskip3truemm\nobreak\noindent	$j=j_1(j_2+2);$

\nobreak\vskip3truemm\nobreak\noindent $
R_{11}=(j+1)(j_2+2)(p_3+j)^{{j_1}-1}[(j_2+1)(p_3+j)^{j_1}+p_2]^{j_2},$

\nobreak\vskip3truemm\nobreak\noindent $
R_{12}=(j+1) p_1(p_3+j)^{{j_1}-1},$

\nobreak\vskip3truemm\nobreak\noindent $ R_{13}=p_1,$

\nobreak\vskip3truemm\nobreak\noindent $
R_{22}=(j+1) (p_3+j)^{{j_1}-1} [(j_2+1)(p_3+j)^{j_1}-{j_2}p_2],$

\nobreak\vskip3truemm\nobreak\noindent $ R_{23}=p_2,$

\nobreak\vskip3truemm\nobreak\noindent $
R_{33}=j+(1-j)p_3,$

\nobreak\vskip3truemm\nobreak\noindent $
A=(1-p_3) [(p_3+j)^{j_1}-p_2]\{[({j_2}+1)(p_3+j)^{j_1}
+p_2]^{j_2+1}-p_1^2\},$

\nobreak\vskip3truemm\nobreak\noindent $ \rho =(j+1)^2(j_2+2) (p_3+j)^
{{j_1}-1}. $

\vskip22truemm\centerline{\bf Class $A8(j_1,j_2)$}
\nobreak\vskip5truemm\nobreak
\noindent $d_A=2 d_1+2 d_2,\quad d_1={s} ({j_1}+1),\quad
d_2={s} ({j_2}+1),\quad d_3=2 {s};$
\nobreak\vskip3truemm\nobreak\noindent $ R_{11}=(j_1+j_2+2)(j_1+1-p_3)^{j_1},$

\nobreak\vskip3truemm\nobreak\noindent $ R_{12}=0,$

\nobreak\vskip3truemm\nobreak\noindent $ R_{13}=-({j_2}+1)p_1,$

\nobreak\vskip3truemm\nobreak\noindent $ R_{22}=(j_1+j_2+2)(j_2+1+p_3)^{j_2},$

\nobreak\vskip3truemm\nobreak\noindent $ R_{23}= ({j_1}+1)p_2,$

\nobreak\vskip3truemm\nobreak\noindent $ R_{33}=({j_1}+1) ({j_2}+1)+({j_1}-
{j_2})p_3,$

\nobreak\vskip3truemm\nobreak\noindent $
A=[(j_1+1-p _3)^{j_1+ 1}-p_1^2][(j_2+1+p_3)^{j_2+1}-p_2^2] ,$

\nobreak\vskip3truemm\nobreak\noindent $ \rho =(j_1+j_2+2)^2. $

\

\vskip22truemm\centerline{\bf Class $A9(j_1)$}
\nobreak\vskip5truemm\nobreak
\noindent $d_A=2 d_1+d_3,\quad d_1=2 {s} ({j_1}+1),\quad
d_2={s} ({j_1}+1),\quad d_3=2 {s};$
\nobreak\vskip3truemm\nobreak\noindent $ R_{11}=(2 j_1+3)(2 j_1+2-p_3)^{2 j_1+
1},$

\nobreak\vskip3truemm\nobreak\noindent $ R_{12}=- p_2 (2 j_1+3)(2 j_1+2-p_3)^
{j_1},$

\nobreak\vskip3truemm\nobreak\noindent $ R_{13}=-p_1,$

\nobreak\vskip3truemm\nobreak\noindent $ R_{22}=(2 j_1+3)(2 j_1+2-p_3)^{j_1},$

\nobreak\vskip3truemm\nobreak\noindent $ R_{23}= -p_2,$

\nobreak\vskip3truemm\nobreak\noindent $ R_{33}=2 j_1+2+(2 j_1+1)p_3,$

\nobreak\vskip3truemm\nobreak\noindent $
A=(1+p _3)[(2 j_1+2-p_3)^{j_1+1}+p_1][(2 j_1+2-p_3)^{j_1+1}-p_1-2 p_2^2],$

\nobreak\vskip3truemm\nobreak\noindent $ \rho =(2 j_1+3)^2 (2
j_1+2-p_3)^{j_1}.$

\vskip22truemm\centerline{\bf Class $A10(j_1)$}
\nobreak\vskip5truemm\nobreak
\noindent $d_A=2 d_1+d_2,\quad d_1={s} ({j_1}+1),\quad
d_2=2 {s} ,\quad d_3= {s};$
\nobreak\vskip3truemm\nobreak\noindent $ R_{11}=(j_1+2)(j_1+1+p_2)^{j_1},$

\nobreak\vskip3truemm\nobreak\noindent $ R_{12}=p_1,$

\nobreak\vskip3truemm\nobreak\noindent $ R_{13}=0,$

\nobreak\vskip3truemm\nobreak\noindent $ R_{22}=j_1+1-j_1 p_2,$

\nobreak\vskip3truemm\nobreak\noindent $ R_{23}= -p_3(j_1+1),$

\nobreak\vskip3truemm\nobreak\noindent $ R_{33}=j_1+2,$

\nobreak\vskip3truemm\nobreak\noindent $
A=[(j_1+1+p _2)^{j_1+ 1}-p_1^2](1-p_2-p_3^2),$

\nobreak\vskip3truemm\nobreak\noindent $ \rho =(j_1+2)^2. $

\

\vfill\eject\vglue10truemm\centerline{\bf Class $B1(j_1)$}
\nobreak\vskip5truemm\nobreak
\noindent $d_A=2 d_1,\quad d_1=6 {s} {j_1},\quad d_2=4 {s},\quad d_3=3 {s};$

\nobreak\vskip3truemm\nobreak\noindent $
R_{11}=[-p_2^2+p_2 p_3^2-2p_3^2+4] [-p_2^3-3 p_2^2+6 p_2 p_3^2-p_3^
4-4  p_3^2+4]^{{j_1}-1},$

\nobreak\vskip3truemm\nobreak\noindent $ R_{12}=0,$

\nobreak\vskip3truemm\nobreak\noindent $ R_{13}=0,$

\nobreak\vskip3truemm\nobreak\noindent $ R_{22}=-p_2+p_3^2+2,$

\nobreak\vskip3truemm\nobreak\noindent $ R_{23}=2 p_3,$

\nobreak\vskip3truemm\nobreak\noindent $ R_{33}=p_2+2,$

\nobreak\vskip3truemm\nobreak\noindent $ A=[-p_2^3-3 p_2^2+6 p_2 p_3^2-p_3^4-4
p_3^2+4]^{j_1}- p_1^2,$

\nobreak\vskip3truemm\nobreak\noindent $\rho =-p_2^2+p_2 p_3^2-2 p_3^2+4.$

\vskip22truemm\centerline{\bf Class $B2$}
\nobreak\vskip5truemm\nobreak
\noindent $d_A=3 d_1,\quad d_1=4 {s},\quad d_2=3 {s} ,\quad d_3=3 {s} ;$

\nobreak\vskip3truemm\nobreak\noindent $ R_{11}=-p_1+p_2^2+p_3^2+2,$

\nobreak\vskip3truemm\nobreak\noindent $ R_{12}=2 p_2,$

\nobreak\vskip3truemm\nobreak\noindent $ R_{13}=2 p_3,$

\nobreak\vskip3truemm\nobreak\noindent $ R_{22}=p_1+2,$

\nobreak\vskip3truemm\nobreak\noindent $ R_{23}=0,$

\nobreak\vskip3truemm\nobreak\noindent $ R_{33}=p_1+2,$

\nobreak\vskip3truemm\nobreak\noindent $
A=-p_1^3-3 p_1^2+6 p_1 (p_2^2+p_3^2)-(p_2^2+p_3^2)^2-4(p_2^2+p_3^2) + 4,$

\nobreak\vskip3truemm\nobreak\noindent $\rho =p_1+2. $

\

\vfill\eject\vglue10truemm\centerline{\bf Class $B3(j_1,j_2)$}
\nobreak\vskip5truemm\nobreak
\noindent $d_A=3 d_1,\quad d_1=2s({j_1}+{j_2}),\quad d_2=3s({j_1}+{j_2})/2,
\quad d_3=s;$

\nobreak\vskip3truemm\nobreak\noindent $j_1\ge j_2$ and $j_1=j_2$ for odd
$d_3$;

\nobreak\vskip3truemm\nobreak\noindent $
R_{11}=[{j_1} {j_2}+({j_1}-{j_2})p_3 ]({j_1}-p_ 3)^{{j_1}-1}
({j_2}+p_3)^{{j_2}-1} [-p_1({j_1}-p_3)^{j_1}({j_2}+p_3)^{j_2}+p _2^2
+2({j_1}-p_3)^{3j_1}({j_2}+p_3)^{3j_2}],$

\nobreak\vskip3truemm\nobreak\noindent $
R_{12}=2 p_2[{j_1} {j_2}+({j_1}-{j_2})p_3 ]
 ({j_1}-p_3)^{2 {j_1}-1}({j_2}+p_3)^{2 {j_2}-1},$

\nobreak\vskip3truemm\nobreak\noindent $ R_{13}=0,$

\nobreak\vskip3truemm\nobreak\noindent $
R_{22}=[{j_1} {j_2}+({j_1}-{j_2})p_3 ]
({j_1}-p_3)^{{j_1}-1}({j_2}+p_3)^{{j_2}-1} [p_1+ 2({j_1}-p_3)^{2j_1}
({j_2}+p_3)^{2j_2}],$

\nobreak\vskip3truemm\nobreak\noindent $ R_{23}=0,$

\nobreak\vskip3truemm\nobreak\noindent $ R_{33}={j_1} {j_2}+({j_1}-{j_2})p_3 ,$

\nobreak\vskip3truemm\nobreak\noindent $
A=-p_1^3 -3 p_1^2({j_1}-p_3)^{2j_1}({j_2}+p_3)^{2j_2}+
6 p_1 p_2^2 ({j_1}-p_3)^{j_1}({ j_2}+p_3)^{j_2}-p_2^4 -4 p_2^2
({j_1}-p_3)^{3j_1}({j_2}+p_3)^{3j_2} +
4 ({j_1}-p_3)^{6j_1}({j_2}+p_3)^{6j_2},$

\nobreak\vskip3truemm\nobreak\noindent $ \rho
=[{j_1} {j_2}+({j_1}-{j_2})p_3 ]^2
 ({j_1}-p_3)^{{j_1} -1}({j_2}+p_3)^{{j_2}-1}. $

\vskip22truemm\centerline{\bf Class $B4(j_1)$}
\nobreak\vskip5truemm\nobreak
\noindent $d_A=3 d_1+d_3,\quad d_1=4 s {j_1} ,\quad d_2=3 s {j_1} ,
\quad d_3=2 s ;$

\nobreak\vskip3truemm\nobreak\noindent $
R_{11}=(6 {j_1}+1)(p_3+6j_1)^{j_1-1}[-p_1 (p_3+6
{j_1})^{j_1}+p_2^2 +2 (p_3+ 6 {j_1})^{3{j_1}}],$

\nobreak\vskip3truemm\nobreak\noindent $ R_{12}=2 (6 {j_1}+1)p_2 (p_3+6 {j_1})^
{2{j_1}-1},$

\nobreak\vskip3truemm\nobreak\noindent $ R_{13}=p_1,$

\nobreak\vskip3truemm\nobreak\noindent $ R_{22}=(6 {j_1}+1)(p_3+6j_1)^{j_1-1}
[p_1+2 (p_3+6 {j_1})^{2{j_1}}],$

\nobreak\vskip3truemm\nobreak\noindent $ R_{23}=p_2,$

\nobreak\vskip3truemm\nobreak\noindent $ R_{33}=6 {j_1}+ (1-6 {j_1})p_3,$

\nobreak\vskip3truemm\nobreak\noindent $
A=(1-p_3) [-p_1^3-3 p_1^2 (p_3+6 {j_1})^{2{j_1}}+6p_1 p_2^2 (p_3+ 6 {j_1})
^{j_1}- p_2^4-4 p_2^2 (p_3+6{j_1})^{3{j_1}}+4(p_3+6j_1)^{6j_1}],$

\nobreak\vskip3truemm\nobreak\noindent $\rho =(6 {j_1}+1)^2(p_3+6 {j_1})^
{{j_1}-1}. $

\vfill\eject\vglue10truemm\centerline{\bf Class $C1(j_1,j_2)$}
\nobreak\vskip5truemm\nobreak
\noindent $d_A=2 d_1,\quad d_1=3 {s} ({j_1}+ 2 {j_2}),\quad d_2=6 {s},
\quad d_3=4 {s};$

\nobreak\vskip3truemm\nobreak\noindent $
R_{11}=\{-p_2^2 [({j_1}-2 {j_2})p_3 +4 (j_1^2-j_1j_2+2 j_2^2)] +4 j_2 p_2
[(3 j_1-j_2)p_3^2 +12j_1 (j_1-j_2) p_3 - 16j_2(j_1^2-j_2^2)]- j_1[j_1p_3^4
+4(j_1^2- 2 j_1 j_2+3 j_2^2) p_3^3 -16j_2 (3 j_1^2-7 j_1j_2-j_2^2)p_3^2 +
192 j_2^2 (j_1^2-j_2^2)p_3 - 256 j_2^3(j_1+j_2)^2]\}[p_2+j_1 (3p_3+4
j_1+4j_2)]^{j_1-1}\{-p_2^2+4j_2p_2(3p_3-4j_2)-(j_1+2j_2)p_3^3+
4j_2[3(j_1-j_2)p_3^2-12j_1j_2p_3+16j_2^2(j_1+j_2)]\}^{j_2-1},$

\nobreak\vskip3truemm\nobreak\noindent $ R_{12}=0,$

\nobreak\vskip3truemm\nobreak\noindent $ R_{13}=0,$

\nobreak\vskip3truemm\nobreak\noindent $
R_{22}=[(-j_1+2j_2)p_3-4j_2(3j_1-2j_2)]p_2+2j_1[(j_1+3j_2)p_3^2-8j_2
(j_1-2j_2)p_3+16j_2^2(j_1+j_2)],$

\nobreak\vskip3truemm\nobreak\noindent $
R_{23}=2(-j_1+2j_2)p_2+j_1 (12j_2-p_3)p_3,$

\nobreak\vskip3truemm\nobreak\noindent $ R_{33}=p_2+2(j_1+j_2)(4j_2-p_3),$

\nobreak\vskip3truemm\nobreak\noindent $
A=[p_2+j_1(3p_3+4j_1+4j_2)]^{j_1}\{-p_2^2+4j_2p_2 (3p_3-4j_2)
-(j_1+2j_2)p_3^3+4j_2[3(j_1-j_2)p_3^2-12j_1j_2p_3+16j_2^2(j_1+j_2)]\}^{j_2}-
p_1^2,$

\nobreak\vskip3truemm\nobreak\noindent $ \rho
=-p_2^2 [ (j_1-2j_2)p_3+4(j_1^2-j_1j_2+2j_2^2)]+4
j_2p_2[(3j_1-j_2)p_3^2 +12j_1(j_1-j_2)p_3 -16j_2
(j_1^2-j_2^2)]\break -j_1[j_1p_3^4+4(j_1^2-2j_1j_2+3j_2^2)
p_3^3-16j_2(3j_1^2-7j_1j_2-j_2^2)p_3^2 +192j_2^2(j_1^2
-j_2^2)p_3  -256 j_2^3 (j_1+j_2)^2].$

\vskip22truemm\centerline{\bf Class $C2(j_1)$}
\nobreak\vskip5truemm\nobreak
\noindent $d_A=2 d_1+d_2,\quad d_1=6 {s} {j_1},\quad d_2=6 {s} ,
\quad d_3=4 {s} ;$

\nobreak\vskip3truemm\nobreak\noindent $
R_{11}= (2j_1+1)(p_3-4j_1)[p_2-2(j_1-1)p_3-8j_1(j_1+1)][-p_2^2+4j_1p_2
(3 p_3-4j_1)- (2j_1+1)p_3^3-12j_1 (j_1-1)p_3^2 -48j_1^2p_3+64j_1^3
(j_1+1)]^{j_1-1},$

\nobreak\vskip3truemm\nobreak\noindent $ R_{12}=-p_1 (4j_1+p_3),$

\nobreak\vskip3truemm\nobreak\noindent $ R_{13}=-2p_1,$

\nobreak\vskip3truemm\nobreak\noindent $
R_{22}=p_2 [(2j_1-1)p_3-4 {j_1} (3-2 {j_1})]+
2(3{j_1}+1)p_3^2+16{j_1}(2{j_1}-1)p_3+32{j_1}^2 ({j_1}+1),$

\nobreak\vskip3truemm\nobreak\noindent $
R_{23}=2(2 j_1-1)p_2+p_3 (12j_1-p_3),$

\nobreak\vskip3truemm\nobreak\noindent $ R_{33}=p_2+2(j_1+1) (4j_1-p_3),$

\nobreak\vskip3truemm\nobreak\noindent $
A=[p_2+3p_3+4({j_1}+1)]\{[-p_2^2+4{j_1}p_2 (3 p_3-4 {j_1})-
(2 {j_1}+1)p_3^3-12 {j_1} ({j_1}-1)p_3^2-48 {j_1}^2p_3+64{j_1}^3
({j_1}+1)]^{j_1}- p_1^2\},$

\nobreak\vskip3truemm\nobreak\noindent $ \rho
=(2j_1+1)(p_3-4{j_1}) [p_2-2(j_1-1)p_3-8j_1(j_1+1)]. $

\

\vfill\eject\vglue10truemm\centerline{\bf Class $C3(j_1)$}
\nobreak\vskip5truemm\nobreak
\noindent $d_A=2 d_1+2 d_2,\quad d_1=3 {s} ({j_1}+1),\quad d_2=6 {s} ,
\quad d_3=4 {s} ;$

\nobreak\vskip3truemm\nobreak\noindent $
R_{11}=({j_1}+3)(4{j_1}+8+p_3)[({j_1}+1) (4 {j_1}+8+3 p_3)+p_2]^{j_1},$

\nobreak\vskip3truemm\nobreak\noindent $ R_{12}=2 p_1 (4+p_3),$

\nobreak\vskip3truemm\nobreak\noindent $ R_{13}=4 p_1,$

\nobreak\vskip3truemm\nobreak\noindent $
R_{22}=p_2[(1-j_1)p_3-4 (3 {j_1}+1)]+2 ({j_1}+1)
[({j_1}+4)p_3^2 -8(j_1-1)p_3+16 ({j_1}+2)],$

\nobreak\vskip3truemm\nobreak\noindent $ R_{23}=2 (1-j_1)p_2+ (j_1+1)p_3
(12-p_3) $

\nobreak\vskip3truemm\nobreak\noindent $ R_{33}=p_2+2 ({j_1}+2) (4-p_3),$

\nobreak\vskip3truemm\nobreak\noindent $
A= [-p_2^2+4p_2(3p_3-4)-(j_1+3)p_3^3+12j_1p_3^2-48(j_1+1)p_3+64(j_1+2
)]\{[(j_1+1)(4j_1+8+3p_3)+p_2]^{j_1+1}-p_1^2\} ,$

\nobreak\vskip3truemm\nobreak\noindent $ \rho = ({j_1}+3)(4 {j_1}+8+p_3). $

\vglue10truemm\centerline{\bf Class $C4$}
\nobreak\vskip5truemm\nobreak
\noindent $d_A=3 d_1,\quad d_1=6 {s},\quad d_2=4 {s} ,\quad d_3=3 {s};$

\nobreak\vskip3truemm\nobreak\noindent $
R_{11}=p_1 (p_2-4)+8 (p_2^2+2 p_2 +8),$

\nobreak\vskip3truemm\nobreak\noindent $ R_{12}=2 p_1-p_2 (p_2-12),$

\nobreak\vskip3truemm\nobreak\noindent $ R_{13}=2 p_3(p_2+4),$

\nobreak\vskip3truemm\nobreak\noindent $ R_{22}=p_1-4 p_2+16,$

\nobreak\vskip3truemm\nobreak\noindent $ R_{23}=4 p_3,$

\nobreak\vskip3truemm\nobreak\noindent $ R_{33}=3(p_2+8),$

\nobreak\vskip3truemm\nobreak\noindent $
A=(p_1+3 p_2-p_3^2+8) [-p_1^2+4p_1(3 p_2-4)-3 p_2^3-48 p_2 +128],$

\nobreak\vskip3truemm\nobreak\noindent $\rho =3(p_2+8). $

\

\vfill\eject\vglue10truemm\centerline{\bf Class $C5(j_1,j_2)$}
\nobreak\vskip5truemm\nobreak
\noindent $d_A=3 d_1,\quad d_1=3 s({j_1}+{j_2}),\quad d_2=2 s ({j_1}+{j_2}),
\quad d_3=s;$

\nobreak\vskip3truemm\nobreak\noindent	$j_1\ge j_2$ and $j_1=j_2$ for odd $d_3
;$

\nobreak\vskip3truemm\nobreak\noindent $
R_{11}=[{j_1} {j_2}+({j_1}-{j_2})p_3]({j_1}-p_3)^{{j_1}-1}
(j_2+p_3)^{{j_2}-1} \{ p_1[p_2-4 (j_1-p_3)^{2j_1} (j_2+p_3)^{2j_2}]+
8 p_2^2 (j_1-p_3)^{j_1}(j_2+p_3)^{j_2} + 16 p_2(j_1-p_3) ^{3j_1}
 (j_2+p_3)^{3j_2}+ 64(j_1-p_3)^{5j_1} (j_2+p_3)^{5j_2}\},$

\nobreak\vskip3truemm\nobreak\noindent $
R_{12}=[{j_1} {j_2}+ ({j_1}-{j_2})p_3] ({j_1}-p_3)^{{j_1}-1}
(j_2+p_3)^{{j_2}-1} [2 p_1(j_1-p_3)^{j_1}(j_2+p_3)^{j_2}-p_2^2+12 p_2
 (j_1-p_3)^{2j_1}(j_2+p_3)^{2j_2}],$

\nobreak\vskip3truemm\nobreak\noindent $ R_{13}=0,$

\nobreak\vskip3truemm\nobreak\noindent $
R_{22}=[ {j_1}{j_2}+({j_1}-{j_2})p_3]
({j_1}-p_3)^{{j_1}-1}(j_2+p_3)^{{j_2}-1} [p_1 -4 p_2(j_1-p_3)^{j_1}
(j_2+p_3)^{j_2} +16 (j_1-p_3)^{3j_1}(j_2+p_3)^{3j_2}] ,$

\nobreak\vskip3truemm\nobreak\noindent $ R_{23}=0,$

\nobreak\vskip3truemm\nobreak\noindent $ R_{33}={j_1} {j_2}+ ({j_1}-{j_2})p_3,$

\nobreak\vskip3truemm\nobreak\noindent $
A=[p_1+3p_2(j_1-p_3)^{j_1}(j_2+p_3)^{j_2}+8(j_1-p_3)^{3j_1}
(j_2+p_3)^{3j_2}]\{-p_1^2+4p_1(j_1-p_3)^{j_1}(j_2+p_3)^{j_2}
[3p_2-4(j_1-p_3)^{2j_1}(j_2+p_3)^{2j_2}] -3p_2^3-16(j_1-p_3)^{4j_1}
(j_2+p_3)^{4j_2}[3p_2-8(j_1-p_3)^{2j_1}(j_2+p_3)^{2j_2}]\}, $

\nobreak\vskip3truemm\nobreak\noindent $ \rho
=[({j_1} {j_2}+ ({j_1}-{j_2})p_3]^2
(j_1-p_3)^{{j_1}-1}(j_2+p_3)^{{j_2}-1} . $

\vskip22truemm\centerline{\bf Class $C6(j_1)$}
\nobreak\vskip5truemm\nobreak
\noindent $d_A=3d_1+ d_3,\quad d_1=6 s {j_1} ,\quad d_2=4 s {j_1} ,
\quad d_3=2 s;$

\nobreak\vskip3truemm\nobreak\noindent $
R_{11}=(9 {j_1}+1) (p_3+9 {j_1})^{j_1-1} \{p_1 [p_2-4(p_3+9 {j_1})^{2{j_1}}]
+8 (p_3+9 {j_1})^{j_1}[p_2^2+ 2 p_2 (p_3+9 {j_1})^{2{j_1}}
 +8 (p_3+9 {j_1})^{4{j_1}}] \},$

\nobreak\vskip3truemm\nobreak\noindent $ R_{12}=(9 {j_1}+1) (p_3+9 {j_1})^
{j_1-1} \{2
p_1 (p_3+9 {j_1})^{j_1} -p_2 [p_2-12 (p_3+9 {j_1})^{2{j_1}}]\},$

\nobreak\vskip3truemm\nobreak\noindent $ R_{13}=p_1,$

\nobreak\vskip3truemm\nobreak\noindent $ R_{22}=(9 {j_1}+1) (p_3+9 {j_1})^
{j_1-1}
\{p_1-4 (p_3+9 {j_1})^{j_1} [p_2-4 (p_3+9 {j_1})^{2{j_1}} ]\};$

\nobreak\vskip3truemm\nobreak\noindent $ R_{23}=p_2,$

\nobreak\vskip3truemm\nobreak\noindent $ R_{33}= 9{j_1}+(1-9 {j_1})p_3,$

\nobreak\vskip3truemm\nobreak\noindent $ A=
(1-p_3)\{p_1+(p_3+9j_1)^{j_1}[3p_2+8(p_3+9j_1)^{2j_1}]\}\{-p_1^2+4p_1
(p_3+ 9j_1)^{j_1}[3p_2-4(p_3+9j_1)^{2j_1}]-3p_2^3-16(p_3+9j_1)^{4j_1}
[3p_2-8 (p_3+9j_1)^ {2j_1}]\},$

\nobreak\vskip3truemm\nobreak\noindent $ \rho =(9 {j_1}+1)^2 (p_3+9 {j_1})^
{j_1-1}.  $

\vfill\eject\vglue10truemm\centerline{\bf Class $D1(j_1)$}
\nobreak\vskip5truemm\nobreak
\noindent $d_A=2 d_1,\quad d_1=15 {s} {j_1},\quad d_2=10 {s},\quad d_3=6 {s};$

\nobreak\vskip3truemm\nobreak\noindent $
R_{11}=[p_2 (p_3-12)+11 p_3^2+24 p_3-576] (4 p_2-p_3^2-8 p_3-192) [-p_ 2^3
+6 p_2^2 (5 p_3-24)-15 p_2 p_3^2 (p_3-12)+p_3^5-55 p_3^4-280 p_3^3-2880
p_3^2+1 10592]^{{j_1}-1},$

\nobreak\vskip3truemm\nobreak\noindent $ R_{12}=0,$

\nobreak\vskip3truemm\nobreak\noindent $ R_{13}=0,$

\nobreak\vskip3truemm\nobreak\noindent $ R_{22}=4 p_2 (p_3-3)-(p_3-48)
(p_3^2+4 p_3+24),$

\nobreak\vskip3truemm\nobreak\noindent $ R_{23}=6 p_2-5 p_3 (p_3-12),$

\nobreak\vskip3truemm\nobreak\noindent $ R_{33}=p_2-2 (7 p_3-48),$

\nobreak\vskip3truemm\nobreak\noindent $
A=[-p_2^3+6 p_2^2 (5 p_3-24)-15 p_2 p_3^2 (p_3-12)+p_3^5-55 p_3^4-280
p_3^3- 2880 p_3^2+110592]^{j_1}-p_1^2,$

\nobreak\vskip3truemm\nobreak\noindent $ \rho =[p_2 (p_3-12)+11 p_3^2+24 p_3
-576]
(4 p_2-p_3^2-8 p_3 - 192).$

\vskip22truemm\centerline{\bf Class $D2$}
\nobreak\vskip5truemm\nobreak
\noindent $d_A=3 d_1,\quad d_1=10 {s},\quad d_2=6 {s} ,\quad d_3=4s;$

\nobreak\vskip3truemm\nobreak\noindent $
R_{11}=p_1 (16p_2-p_3^2-192)-4p_2^3+704 p_2^2+2 p_2 (p_3^3-156
p_3^2+5376) +46  p_3^4+192 p_3^3-8064 p_3^2+294912,$

\nobreak\vskip3truemm\nobreak\noindent $
R_{12}=24 p_1-20 p_2^2+5 p_2 (p_3^2+192)+2 p_3^2 (p_3-96),$

\nobreak\vskip3truemm\nobreak\noindent $ R_{13}=p_3 (12 p_2-p_3^2),$

\nobreak\vskip3truemm\nobreak\noindent $ R_{22}=p_1-2 (28 p_2-11 p_3^2-768),$

\nobreak\vskip3truemm\nobreak\noindent $ R_{23}=p_3 (p_3+48),$

\nobreak\vskip3truemm\nobreak\noindent $ R_{33}=p_2+6 p_3+96,$

\nobreak\vskip3truemm\nobreak\noindent $
A=-p_1^3+3 p_1^2 (40 p_2-5 p_3^2-768)-3 p_1 [20p_2^3-5 p_2^2 (p_3^2+192)-
10 p_2 p_3^2 (p_3-48)+4 p_3^4 (p_3-20)]+4p_2^5-880p_2^4-10 p_2^3 (p_3^3-84
p_3^2 +1792)-5 p_2^2 (63 p_3^4+128p_3^3-5952p_3^2+147456)+40 p_2 p_3^2 (
p_3^4+30p_3^3-456 p_3^2-768 p_3+18432)-4(45 p_3^7-580 p_3^6+2160 p_3^5-
20160p_3^4-184320 p_3^3+4423680 p_3^2-113246208),$

\nobreak\vskip3truemm\nobreak\noindent $ \rho =p_2+6 p_3+96. $

\vfill\eject\vglue8truemm\centerline{\bf Class $D3(j_1,j_2)$}
\nobreak\vskip5truemm\nobreak
\noindent $d_A=3 d_1,\quad d_1=5s({j_1}+{j_2}),\quad d_2=3s({j_1}+{j_2}),\quad
d_3=s;$

\nobreak\vskip3truemm\nobreak\noindent	$j_1\ge j_2$ and $j_2=j_1$ for odd $d_3
;$

\nobreak\vskip3truemm\nobreak\noindent $
R_{11}=[{j_1}{j_2}+(j_1-j_2)p_3]({j_1}-p_ 3)^{{j_1}-1}(j_2+p_3)^{{j_2}-1}
\{4 p_1 ({j_1}-p_3)^{j_1}(j_2+p_3)^{j_2}[ p_2-3 ({j_1}-p_3)^{3j_1}
(j_2+p_3)^{3j_2}]\break
-[p_2-48 ({j_1}-p_3)^{3j_1}(j_2+p_3)^{3j_2}][ p_2^2+4 p_2
 ({j_1}-p_3)^{3j_1}(j_2+p_3)^{3j_2}+24 ({j_1}-p_3)^{6j_1}(j_2+p_3)^{6j_2}]\},$

\nobreak\vskip3truemm\nobreak\noindent $
R_{12}=[{j_1}{j_2}+(j_1-j_2)p_3]({j_1 }-p_3)^{2 {j_1}-1}
(j_2+p_3)^{2 {j_2}-1}[6 p_1 ({j_1}-p_3)^{j_1}(j_2+p_3)^{j_2}
-5 p_2^2+60 p_2({j_1}-p_3)^{3j_1}(j_2+p_3)^{3j_2} ] ,$

\nobreak\vskip3truemm\nobreak\noindent $ R_{13}=0,$

\nobreak\vskip3truemm\nobreak\noindent $
R_{22}=[{j_1}{j_2}+(j_1-j_2)p_3]({j_1}-p_3)^{{j_1}-1}(j_2+p_3)^{{j_2}-1}
[p_1-14 p_2 ({j_1}-p_3) ^{2j_1}(j_2+p_3)^{2j_2}
+96 ({j_1}-p_3)^{5j_1}(j_2+p_3)^{5j_2}],$

\nobreak\vskip3truemm\nobreak\noindent $ R_{23}=0,$

\nobreak\vskip3truemm\nobreak\noindent $ R_{33}={j_1} {j_2}+({j_1}-{j_2})p_3,$

\nobreak\vskip3truemm\nobreak\noindent $
A=-p_1^3+6 p_1^2 ({j_1}-p_3)^{2j_1}(j_2+p_3)^{2j_2}[5p_2
-24({j_1}-p_3)^{3j_1}(j_2+p_3)^{3j_2}]
-15 p_1 p_2^2 ({j_1}-p_3)^{j_1}(j_2+p_3)^ {j_2} [p_2-
12({j_1}-p_3)^{3j_1}(j_2+p_3)^{3j_2}]
+p_2^5-55 p_2^4 ({j_1}-p_3)^{3j_1}(j_2+p_3)^{3j_2}
-280 p_2^3 ({j_1}-p_3)^{6j_1}(j_2+p_3)^{6j_2}
-2880 p_2^2 ({j_1}-p_3)^{9j_1}(j_2+p_3)^{9j_2}
+110592 ({j_1}-p_3)^{15j_1}(j_2+p_3)^{15j_2},$

\nobreak\vskip3truemm\nobreak\noindent $ \rho
=[{j_1}{j_2}+(j_1-j_2)p_3]^2({j_1}-p_3)^{{j_1}-1}(j_2+p_3)^{{j_2}-1}.$

\vskip12truemm\centerline{\bf Class $D4(j_1)$}
\nobreak\vskip5truemm\nobreak
\noindent $d_A=3 d_1+d_3, \quad d_1=10 s {j_1} ,\quad d_2=6 sj_1 ,\quad
d_3=2 s;$

\nobreak\vskip3truemm\nobreak\noindent $
R_{11}=(15{j_1}+1)(p_3+15j_1)^{{j_1}-1}
\{4p_1 (p_3+15j_1)^{j_1} [p_2-3(p_3+15j_1)^{3j_1}]-
[p_2-48(p_3+15j_1)^{3j_1}]\break
[p_2^2+4 p_2(p_3+15j_1)^{3j_1}+ 24(p_3+15j_1)^{6j_1}]\},$

\nobreak\vskip3truemm\nobreak\noindent $
R_{12}=(15{j_1}+1)(p_3+15j_1)^{2{j_1}-1}
[6p_1(p_3+15j_1)^{j_1}-5p_2^2+60p_2(p_3+15j_1)^{3j_1}],$

\nobreak\vskip3truemm\nobreak\noindent $
R_{13}=p_1,$

\nobreak\vskip3truemm\nobreak\noindent $
R_{22}=(15{j_1}+1)(p_3+15j_1)^{{j_1}-1}
[p_1-14p_2(p_3+15j_1)^{2j_1}+96(p_3+15j_1)^{5j_1}],$

\nobreak\vskip3truemm\nobreak\noindent $
R_{23}=p_2,$

\nobreak\vskip3truemm\nobreak\noindent $
R_{33}=15{j_1}+(1-15{j_1})p_3,$

\nobreak\vskip3truemm\nobreak\noindent $
A=(1-p_3)[-p_1^3+6p_1^2(p_3+15j_1)^{2j_1}[5p_2-24(p_3+15j_1)^{3j_1}]
-15p_1p_2^2(p_3+15j_1)^{j_1}[p_2-12(p_3+15j_1)^{3j_1}]
+p_2^5-55p_2^4(p_3+15j_1)^{3j_1}-280p_2^3(p_3+15j_1)^{6j_1}
-2880p_2^2(p_3+15j_1)^{9j_1}+110592(p_3+15j_1)^{15j_1}],$

\nobreak\vskip3truemm\nobreak\noindent $\rho =(15{j_1}+1)^2(p_3+15j_1)^{{j_1}-1
}.$

\vfill\eject\vglue10truemm\centerline{\bf Class $E1$}
\nobreak\vskip5truemm\nobreak
\noindent $d_A=4 d_1,\quad d_1=5 {s},\quad d_2=4 {s} ,\quad d_3=3 {s};$

\nobreak\vskip3truemm\nobreak\noindent $
R_{11}=-4 p_1 p_3+3p_2^2-27 p_2+18 (2 p_3^2+9),$

\nobreak\vskip3truemm\nobreak\noindent $ R_{12}=-6p_1+p_3 (5 p_2+54),$

\nobreak\vskip3truemm\nobreak\noindent $ R_{13}=2 (3 p_2+p_3^2),$

\nobreak\vskip3truemm\nobreak\noindent $ R_{22}=3 p_2+2 (4 p_3^2+27),$

\nobreak\vskip3truemm\nobreak\noindent $ R_{23}=p_1+12 p_3,$

\nobreak\vskip3truemm\nobreak\noindent $ R_{33}=p_2+9,$

\nobreak\vskip3truemm\nobreak\noindent $
A= p_1^4+40 p_1^3 p_3-6p_1^2 [5 p_2^2+5 p_2 (p_3^2+9)-30 p_3^2+162]
+4 p_1 p_3 [5 p_2^3+45 p_2^2-30 p_2 (p_3^2-27)+4 p_3^2 (4
p_3^2-135)]-3p_2^5+45 p_2^4-5
 p_2^3 (2 p_3^2-27)-15 p_2^2 (p_3^4+162 p_3^2+243)+1620 p_2 p_3^2
(p_3^2+6)-4(80	p_3^6+135 p_3^4+7290 p_3^2-19683),$

\nobreak\vskip3truemm\nobreak\noindent $\rho =1. $

\vskip22truemm\centerline{\bf Class $E2$}
\nobreak\vskip3truemm\nobreak
\noindent $d_A=4 d_1,\quad d_1=6 {s},\quad d_2=4 {s} ,\quad d_3=4 {s};$

\nobreak\vskip3truemm\nobreak\noindent $ R_{11}=-4p_1+5 p_2^2+5 p_3^2+12,$

\nobreak\vskip3truemm\nobreak\noindent $ R_{12}=2 p_2 (p_3+3),$

\nobreak\vskip3truemm\nobreak\noindent $ R_{13}=p_2^2-p_3 (p_3-6),$

\nobreak\vskip3truemm\nobreak\noindent $ R_{22}=p_1+3 p_3+6,$

\nobreak\vskip3truemm\nobreak\noindent $ R_{23}=3 p_2,$

\nobreak\vskip3truemm\nobreak\noindent $ R_{33}=p_1-3 p_3+6,$

\nobreak\vskip3truemm\nobreak\noindent $
A=-p_1^4-16 p_1^3+2 p_1^2 [3 p_2^2 (p_3+5)-p_3^3+15 p_3^2-36]-12 p_1
[p_2^4+2 p_2^2 (p_3^2+6 p_3-6)+p_3^2 (p_3^2-4 p_3-12)]+p_2^6-3 p_2^4 (2
p_3^2-18 p_3 +15)+9 p_2^2 (p_3^4+4 p_3^3-10 p_3^2+24 p_3-24)-9 (p_3+2)^2
(2p_3^3-3p_3^2+12 p_3-12),$

\nobreak\vskip3truemm\nobreak\noindent $\rho =1. $

\vfill\eject\vglue10truemm\centerline{\bf Class $E3$}
\nobreak\vskip5truemm\nobreak
\noindent $d_A=4 d_1,\quad d_1=8s,\quad d_2=6 {s} ,\quad d_3=4 {s};$

\nobreak\vskip3truemm\nobreak\noindent $
R_{11}=2 p_1 (p_2-54)+15 p_2^2+216 p_2 p_3+324(4p_3^2+18 p_3+81),$

\nobreak\vskip3truemm\nobreak\noindent $
R_{12}=6 p_1 p_3-p_2^2+18 p_2 (p_3+12)+1620 p_3,$

\nobreak\vskip3truemm\nobreak\noindent $ R_{13}=6 p_1-p_2 (p_3-27)+54 p_3,$

\nobreak\vskip3truemm\nobreak\noindent $
R_{22}=4 [3 p_1-p_2 (p_3+9)+27(p_3^2-2p_3+27)],$

\nobreak\vskip3truemm\nobreak\noindent $ R_{23}=p_1-3 p_2-6 p_3^2+108 p_3,$

\nobreak\vskip3truemm\nobreak\noindent $ R_{33}=p_2-12 p_3+81,$

\nobreak\vskip3truemm\nobreak\noindent $
A=(p_1+4 p_2+36 p_3+162)\{p_1^3-6 p_1^2 (6 p_2+4 p_3^2-90 p_3+243)
+12p_1[2 p_2^2 (p_3+9)-12 p_2 (2 p_3^2+18 p_3-81)+3 (4 p_3^4+24 p_3^3-540
p_3^2+2916 p _3-6561)]
-4 [p_2^4+4 p_2^2 (2 p_3^3-54 p_3^2+729 p_3+729)+36 p_2 (4 p_3^4-120 p_3^3-291
6 p_3+ 6561)+162 (8 p_3^5-12 p_3^4+288 p_3^3+13122 p_3-59049)]\},$

\nobreak\vskip3truemm\nobreak\noindent $ \rho =1. $

\vskip22truemm\centerline{\bf Class $E4$}
\nobreak\vskip5truemm\nobreak
\noindent $d_A=4 d_1,\quad d_1=12 s,\quad d_2=8 s ,\quad d_3=6 s ;$

\nobreak\vskip3truemm\nobreak\noindent $
R_{11}=-6 [12 p_1 p_3-21 p_2^2-p_2 (11 p_3^2-864)-36 (17 p_3^2+97
2)],$

\nobreak\vskip3truemm\nobreak\noindent $ R_{12}=-6 [12 p_1-p_3 (21 p_2+p_3^2+
756)],$

\nobreak\vskip3truemm\nobreak\noindent $ R_{13}=p_2^2+180 p_2+60 p_3^2,$

\nobreak\vskip3truemm\nobreak\noindent $ R_{22}=6(9 p_2+7 p_3^2+972),$

\nobreak\vskip3truemm\nobreak\noindent $ R_{23}=p_1+180 p_3,$

\nobreak\vskip3truemm\nobreak\noindent $ R_{33}=5 p_2+324,$

\nobreak\vskip3truemm\nobreak\noindent $
A=p_1^4+576 p_1^3 p_3-2 p_1^2 [p_2^3+810 p_2^2+36 p_2 (11 p_3^2+1944)+6
(p_3^4-4536 p_3^2+314928)]+288 p_1 p_3 [7 p_2^3+p_2^2 (p_3^2+648)-144 p_2
(2 p_3^2-243) +20 p_3^2 (p_3^2-324)]+p_2^6-324 p_2^5-36 p_2^4 (14 p_3^2-729)
-12 p_2^3 (p_3^4+ 3780 p_3^2-104976)+
216 p_2^2 (95 p_3^4-40824 p_3^2-944784)-2160 p_2 p_3^2 (p_3^4-1080 p_3^2-
104976)+36 (p_3^8-3696 p_3^6-536544 p_3^4-136048896 p_3^2+11019960576),$

\nobreak\vskip3truemm\nobreak\noindent $ \rho =1. $

\vfill\eject\vglue10truemm\centerline{\bf Class $E5$}
\nobreak\vskip5truemm\nobreak
\noindent $d_A=4 d_1,\quad d_1=30 {s},\quad d_2=20 {s} ,\quad d_3=12 {s};$

\nobreak\vskip3truemm\nobreak\noindent $
R_{11}=90 p_1 (12 p_2+p_3^2+2^3\cdot 3^3\cdot 5\cdot p_3-2^6\cdot 3^6\cdot
5^2)-36 p_2^2 (19 p_3-2^2\cdot 3^3\cdot 5\cdot 47) - p_2 (29 p_3^3-
2^4\cdot 3^4\cdot 5\cdot p_3^2-2^7\cdot 3^7\cdot 5^2\cdot 7^2\cdot p_3-2^{11}
\cdot 3^{10}\cdot 5^4)+45 (11\cdot 19\cdot p_3^4-2^4\cdot 3^3\cdot 5\cdot
157\cdot p_3^3+2^6\cdot 3^9\cdot 5^3\cdot 7\cdot p_3^2+2^{13}\cdot 3^{11}\cdot
5^4\cdot p_3+2^{12}\cdot 3^{15}\cdot 5^6) ,$

\nobreak\vskip3truemm\nobreak\noindent $
R_{12}=360 p_1
(p_3+2^2\cdot 3^3\cdot 5)-486p_2^2-9p_2(19p_3^2-2^3\cdot 3^4\cdot 5\cdot p_3
-2^7\cdot 3^7\cdot 5^3)-p_3(p_3^3-2^4\cdot 3^4\cdot 5\cdot 7\cdot p_3^2
+2^3\cdot 3^7\cdot 5^3\cdot 13\cdot p_3-2^7\cdot 3^{10}\cdot 5^4\cdot 19),$

\nobreak\vskip3truemm\nobreak\noindent $
R_{13}= 2160 p_1+p_2^2-1980 p_2 (p_3-2^3\cdot 3^3\cdot 5)-55 p_3^2(p_3-2^2\cdot
3^5\cdot 5),$

\nobreak\vskip3truemm\nobreak\noindent $
R_{22}=
540 p_1-324 p_2 (p_3+3^4\cdot 5^2)-19 p_3^3+
2^2\cdot 3^7\cdot 5^2\cdot p_3^2-2^3\cdot 3^8\cdot 5^3\cdot 7\cdot p_3 +
2^7\cdot 3^{12}\cdot 5^5,$

\nobreak\vskip3truemm\nobreak\noindent $ R_{23}=p_1-810 p_2-495 p_3 (p_3-2^2
\cdot 3^3\cdot 5^2),$

\nobreak\vskip3truemm\nobreak\noindent $ R_{33}=11 p_2-6750(p_3-2^4\cdot 3^4),$

\nobreak\vskip3truemm\nobreak\noindent
$ A=p_1^4- 360p_1^3[36p_2+5(p_3^2-2^3\cdot 3^3\cdot 5^2\cdot p_3+2^6\cdot
3^7\cdot 5^2)]+ 2p_1^2[-p_2^3+810p_2^2(13p_3+2^3\cdot 3^3\cdot 5\cdot
11)+15p_2 (29p_3^3-2^2\cdot 3^5\cdot 5^2\cdot 19\cdot p_3^2
-2^7\cdot 3^9\cdot 5^3\cdot p_3+2^{10}\cdot 3^{12}\cdot 5^4)
+p_3^5+3^3\cdot 5^3\cdot 53\cdot p_3^4+
2^3\cdot 3^7\cdot 5^4\cdot 53\cdot p_3^3-2^6\cdot 3^{11}\cdot 5^7\cdot
p_3^2+2^{11}\cdot 3^{15}\cdot 5^7\cdot p_3-2^{12}\cdot 3^{19}\cdot 5^8]
-8p_1[2268p_2^4+45p_2^3(19p_3^2+2^5\cdot 3^3\cdot 5\cdot p_3
-2^6\cdot 3^8\cdot 5^2)+15p_2^2(p_3^4-2\cdot 3^3\cdot 5\cdot 137\cdot p_3^3-
2^7\cdot 3^8\cdot5^3\cdot p_3^2+2^7\cdot 3^{11}\cdot 5^4 \cdot 11\cdot p_3-
2^{11}\cdot3^{15}\cdot 5^5)+135p_2p_3(11\cdot 17\cdot p_3^4+2^2\cdot 3^3\cdot
5^3\cdot97\cdot p_3^3-2^7\cdot 3^8\cdot 5^5\cdot p_3 ^2+2^8\cdot 3^{13}\cdot
5^6\cdot p_3-2^{12}\cdot 3^{15}\cdot 5^7)-275p_3^3(p_3^4+2\cdot 3^5\cdot 5^2
\cdot p_3^3-2^3\cdot 3^8\cdot 5^2\cdot 19 \cdot p_3^2+2^6\cdot 3^{12}\cdot 5^5
\cdot p_3-2^9\cdot 3^{15}\cdot 5^6)] +p_2^6+972p_2^5(5p_3-2^3\cdot 3^4\cdot 11
 )+30p_2^4( 19p_3^3-2\cdot 3^6\cdot 23\cdot p_3^2
-2^5\cdot 3^8\cdot 5^2\cdot p_3-2^7\cdot 3^{12}\cdot
5^5)+2p_2^3(p_3^5-3^5\cdot 5\cdot 107\cdot p_3^4
-2^5\cdot 3^7\cdot 5^4\cdot 37\cdot p_3^3+2^{10}\cdot 3^{10}\cdot 5^4\cdot
113\cdot p_3^2-2^{14}\cdot 3^{15}\cdot 5^6\cdot p_3-2^{12}\cdot 3^{20}\cdot
5^7)+15 p_2^2(11\cdot 109\cdot p_3^6+2^2\cdot 3^5\cdot 5\cdot 11\cdot 463\cdot
p_3^5-2^4\cdot 3^8\cdot 5^3\cdot 11\cdot 563\cdot p_3^4+2^{11}\cdot
3^{12}\cdot 5^5 \cdot 59\cdot p_3^3-2^{11}\cdot 3^{16}\cdot 5^6\cdot 41\cdot
p_3^2-2^{14}\cdot 3^{20}\cdot 5^8\cdot p_3-2^{17}\cdot 3^{24}\cdot
5^9)-330p_2p_3^2(p_3 ^6+3^4\cdot 5^2\cdot 17\cdot p_3^5-2^2\cdot 3^7\cdot
5^3\cdot 107\cdot p_3^4+2^5\cdot 3^{10}\cdot 5^4\cdot 499\cdot p_3^3-2^8\cdot
3^{15}\cdot 5^6\cdot 7\cdot p_ 3^2-2^{12}\cdot 3^{18}\cdot 5^8\cdot
p_3-2^{14}\cdot 3^{22}\cdot 5^9)+p_3^{10}+2\cdot 5^3\cdot 17\cdot 23\cdot
p_3^9-3^5\cdot 5^4\cdot 1913\cdot p_3^8+2^4\cdot 3^{13}\cdot 5^6\cdot 7\cdot
p_3^7-2^6\cdot 3^{13}\cdot 5^7\cdot 17\cdot 47\cdot p_3^6-2^{10}\cdot
3^{17}\cdot 5^8\cdot 13\cdot 23\cdot p_3^5-2^{12}\cdot 3^{20}\cdot 5^{10}\cdot
173\cdot p_3^4-2^{16}\cdot 3^{24}\cdot 5^{13}\cdot p_3^3-2^{19}\cdot
3^{29}\cdot 5^{13}\cdot p_3^2+2^{24}\cdot 3^{36}\cdot 5^{15}.$

\nobreak\vskip3truemm\nobreak\noindent $ \rho =1. $

\vskip3truecm
In order to allow the reader to get a glimmering of the shape of possible
4-dimensional orbit spaces, we conclude by giving a graphical representation
of the (intersections with the coordinate planes of) the positivity regions
$\bar{\cal S}$ of some of the allowable $\hat P$-matrices listed above.  Our
examples are selected by picking up the solution corresponding to the lowest
values of the parameters $j$ and $s$ in each tower of allowable solutions, but
for the case A(1,2,2).	The sections $p_1=0$, $p_2=0$ and $p_3=0$ of
$\bar{\cal S}$ will be represented in three distinct graphs on the same row,
with axes $(p_3,p_2)$, $(p_3,p_1)$ and, respectively $(p_2,p_1)$.  The unit
lengths in the two coordinate axes of a graph, or in the homonymous axes of
graphs lying in different rows, may be different, but they they have been
chosen to be the same in homonymous axes of graphs lying in the same row.

\vfill\eject
\section{Acknowledgments}

This paper is partially supported by INFN and MURST 60\%.

\end{document}